# Observation of Gapped Topological Surface States and Isolated Surface Resonances in PdTe$_2$ Ultrathin Films

Jacob Cook,[1] Sougata Mardanya,[2] Qiangsheng Lu,[1] Clayton Conner,[1] James McMillen,[1] Chi Chen,[1] Mathew Snyder,[1] Xiaoqian Zhang,[1] Tay-Rong Chang,[2*] Guang Bian[1*]

*Correspondence and requests for materials should be addressed to G.B. (E-mail: biang@missouri.edu) and T.-R.C. (E-mail: u32trc00@phys.ncku.edu.tw)

[1]Department of Physics and Astronomy, University of Missouri, Columbia, Missouri 65211, USA

[2]Department of Physics, National Cheng Kung University, Tainan 701, Taiwan

## Abstract

The superconductor PdTe$_2$ is known to host bulk Dirac bands and topological surface states. The coexistence of superconductivity and topological surface states makes PdTe$_2$ a promising platform for exploring topological superconductivity and Majorana bound states. In this work, we report the layer-by-layer molecular beam epitaxy growth and spectroscopic characterization of high quality  PdTe$_2$ films with thickness down to 3 monolayers (ML). In the 3 ML PdTe$_2$ film, we observed spin-polarized surface resonance states, which are isolated from the bulk bands due to the quantum size effects.  In addition, the hybridization of surface states on opposite faces leads to a thickness-dependent gap in the topological surface Dirac bands. Our photoemission results show clearly that the size of the hybridization gap increases as the film thickness is reduced. The success in growing high quality PdTe$_2$ films by state-of-art molecular beam epitaxy technique and the observation of surface resonances and gaped topological surface states sheds light on the applications of PdTe$_2$ quantum films in spintronics and topological computation.

The reduction of bulk crystal into two-dimensional (2D) while maintaining high crystallinity is of the utmost importance for studying interfacial effects, the behavior of exotic properties in reduced dimensionality, and their practical incorporation in electronic devices. One of the most extensively sought-after phenomena in condensed matter is topological superconductivity (TSC) due to the possibility of such novel electronic states hosting massless Majorana fermions, which arise from the spin-triplet *p*-wave superconductivity, and surface electronic states[1,2]. These exotic quasiparticles are of significant interest for their application in the field of quantum computing, due to the topologically protected nature of Majorana bound states which is robust against decoherence[3-5].

Efforts to realize topological superconductivity have taken multiple different approaches, including inducing superconductivity in topological insulators via proximity effect with a *s*-wave superconductor[6], extrinsic doping of topological materials[7-9], and subjecting topological materials to extreme pressures[10]. However, there are many technical obstacles in those approaches. For example, the proximity effect heavily relies on the interface conditions while doping materials can lead to uncontrolled surface defects and reduced uniformity. Considering the limitations on the existing approaches to topological superconductivity, we are motivated to search for materials that can intrinsically host both topologically non-trivial surface states (TSS), and superconductivity[11,12]. Transition metal dichalcogenides (TMD) have recently become of great interest in this effort because of their connection to superconductivity and topologically non-trivial electronic band structures[13,14].

The semimetallic TMD $PdTe_2$ has been known to be a bulk superconductor with a transition temperature of 1.7 K. Over the past 5 years, it has gained ever increasing attention due to the discovery of intrinsic type-I superconductivity[15-19] with anomalous type-II surface behavior[20,21], a Type-II Dirac cone[22-25], topologically non-trivial surface states[22,26,27], and a semiconductor to conductor transition from 1 to 2 monolayers (ML)[28]. The coexistence of the topological surface states and superconductivity suggests that $PdTe_2$ is a candidate to host TSC. Compared with the bulk materials, thin films offer technical advantages including band engineering by quantum size effects and construction of functional heterostructures. Previously, thin films of $PdTe_2$ have been grown on bilayer graphene (BLG)/SiC substrates[29], but have only formed films thicker than 6ML, well above the 2D limit. $PdTe_2$ monolayers have been grown on $SrTiO_3$[28], however, such heterostructures do not allow for experimental analysis of the band structure.

Here, we perform the layer-by-layer epitaxial growth of high quality PdTe$_2$ films on graphene (Gr)/SiC and capture the angle-resolved photoemission spectroscopy (ARPES) image of the band structure in the 2D limit. The structural quality of these ultrathin films was verified using *in-situ* scanning tunneling spectroscopy (STM) and the ARPES measurements. The pairing of ARPES results and first-principles calculation show the two-dimensionality of the crystal and reveal a thickness-dependent gap opening at multiple topologically non-trivial surface bands, including ones near the Fermi level. These states could be potentially used to study the interplay of intrinsic superconductivity and topologically nontrivial electronic states.

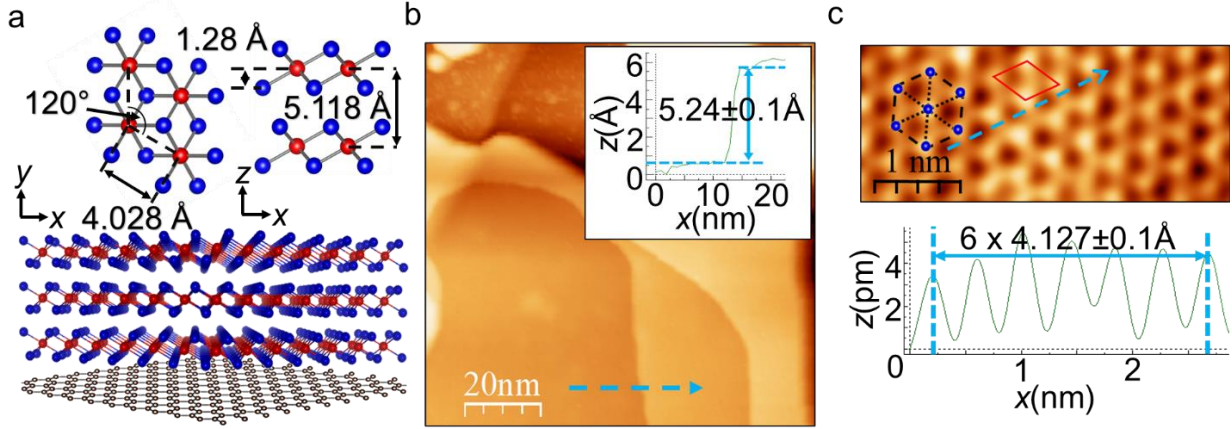

**Figure 1.** Crystal structures and STM measurements of PdTe$_2$. (a) The lattice parameters and crystal structure of PdTe$_2$ crystals, shown with a graphene layer beneath. (b) STM topography of PdTe$_2$ grown on Gr/SiC substrate. Inset shows the height profile designated by the blue arrow. The step height is $5.24 \pm 0.1$Å, slightly larger than the bulk out-of-plane lattice constant $c = 5.118$Å. (c) Atomic-resolution STM image of the PdTe$_2$ surface showing a trigonal lattice. Lattice periodicity is determined by a cut along the blue arrow, shown below the image.

PdTe$_2$ is a layered trigonal crystal structure with a unit cell of a hexagonal Pd layer sandwiched between two Te layers. Bulk PdTe$_2$ has the 1T-CdI$_2$ structure type and a P$\bar{3}$m1 space group with AA stacking between layers. The bulk lattice parameters are $a = 4.028$Å in-plane and $c = 5.118$Å out of plane (Figure 1a). We grew PdTe$_2$ thin films on the Gr/SiC surface at temperatures ranging from 190°C to 220°C. The optimum crystal quality was found at 210°C. The STM topography of the sample (Figure 1b) shows high quality PdTe$_2$ crystals on the Gr surface as thin as 3 monolayers (ML). The step between 3ML and 4ML growths shows $c = 5.24\pm0.1$Å (Figure 1b), which has an expansion of about 2.3% compared to the bulk value. Atomic resolution STM images of the 3ML crystal surface (Figure 1c) show a trigonal lattice structure which arises from the surface Te atoms. This gives us $a = 4.127\pm0.1$ Å (in-plane lattice constant), an expansion of 2.1% compared to the bulk value. The orientation of the hexagonal PdTe$_2$ crystal is shown to be aligned with the honeycomb lattice of graphene, which means the Γ-K direction of graphene is parallel to the Γ-K direction of PdTe$_2$ in reciprocal space (see Supplementary Information).

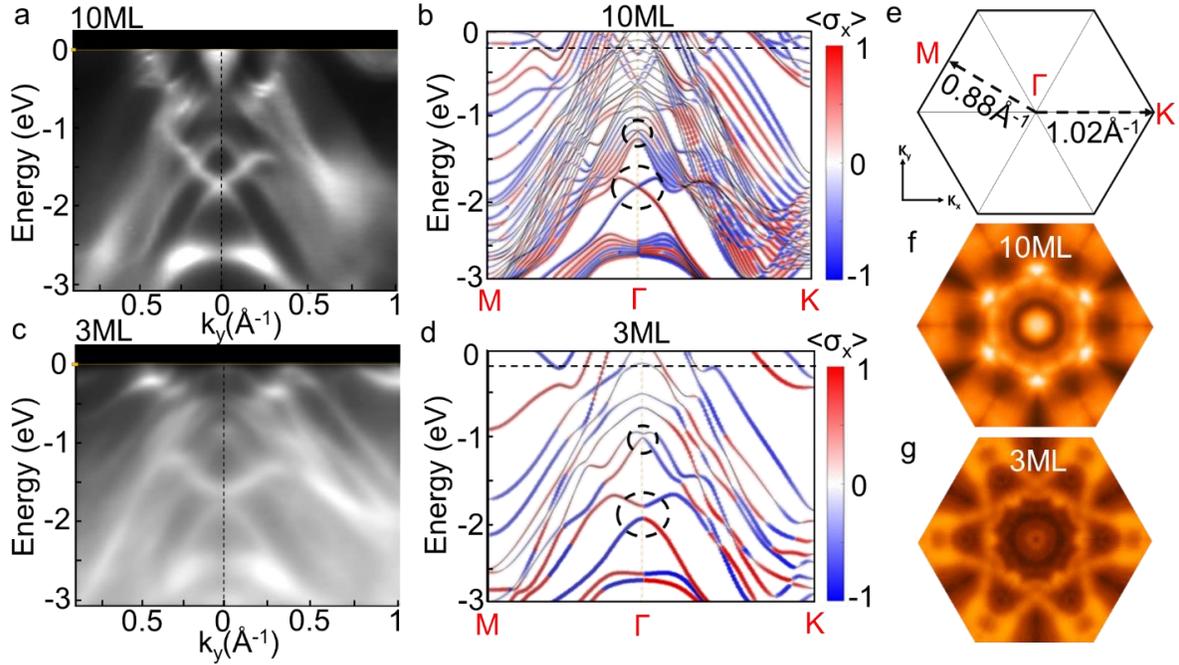

**Figure 2.** Low-temperature ARPES and spin-resolved DFT calculations of PdTe$_2$ films. ARPES spectrum along the M-Γ-K high symmetry direction with corresponding spin-resolved DFT calculations for (a, b) 10ML and (c, d) 3ML. The lower black circle marks the topological Dirac cone at -1.67 eV with a gap opening for 3ML, while the upper one marks a spin-polarized Rashba resonance at -1.2 eV. (e) The surface Brillouin zone (SBZ) of PdTe$_2$, showing the Γ-M and Γ-K high symmetry directions. (f, g) The fermi-level iso-energy contour of the first BZ for (f) 10ML and (g) 3ML.

Using ARPES, we mapped the electronic structure of films with thicknesses varying from 3ML, 4ML, 5ML, 7ML, and 10ML. DFT calculations were performed along the M-Γ-K high symmetry directions for each film thickness as well as for the 1ML and 2ML cases (see Supplementary Information). When comparing the 10ML results (Figure 2a, b) along M-Γ-K, the known topological surface Dirac cone at 1.67eV below the Fermi level is very prominent as well as the large electron pocket around $\bar{\Gamma}$ near the Fermi surface. The type-II bulk Dirac node can be found at $\bar{\Gamma}$ 0.7 eV below the fermi level where a set of electron-like quantum well subbands (QWSs) meet a set of hole-like QWSs. The spin resolved DFT bands of 10ML (Figure 2b) verifies the helical spin texture of the topological surface Dirac cone and shows another pair of Rashba split bands at a higher binding energy of about 1.2 eV at $\bar{\Gamma}$ (marked by a dashed circle in Fig. 2b). This pair of bands are spin polarized and lie in the bulk band region, suggesting a surface resonance nature[14]. As the film thickness decreases to 3ML (Figure 2c, d), the band structure shows the topological surface Dirac cone opens a gap, visible in both the DFT and ARPES data. This gap is due to the hybridization effect between the surface states on opposite surfaces of ultrathin films. The spin resolved DFT bands of 3ML (Figure 2d) also show a small gap opening at the band crossing of the Rashba surface resonance bands at $\bar{\Gamma}$ 1.02 eV below the Fermi level.

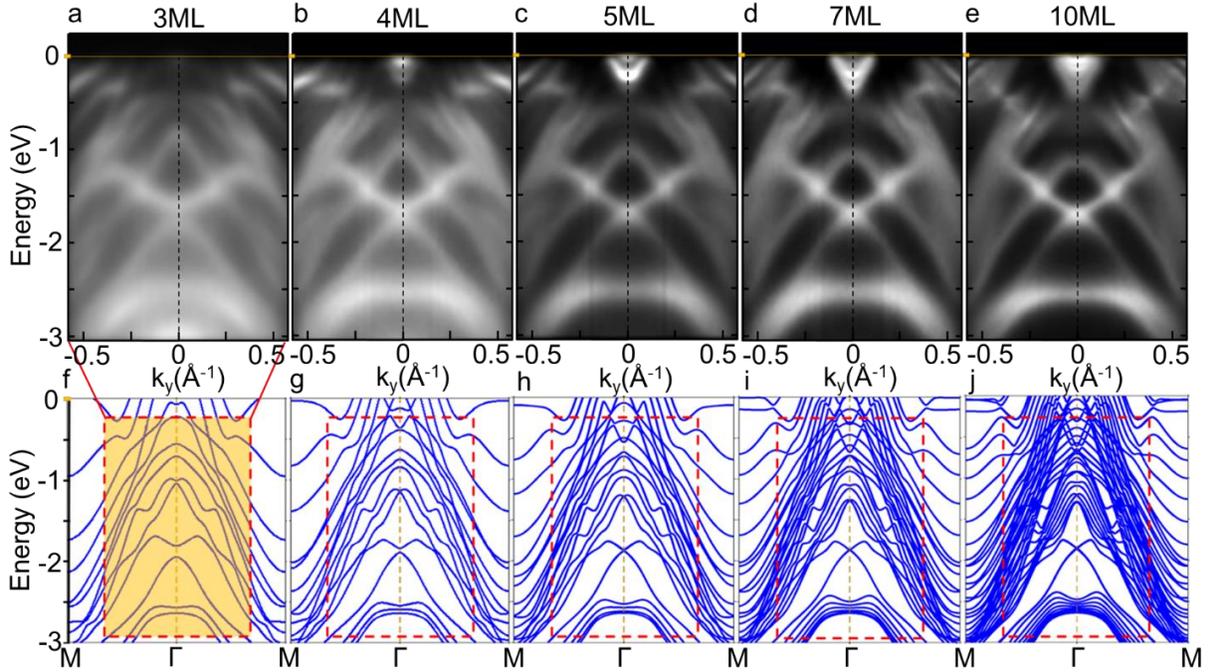

**Figure 3.** Thickness-dependent ARPES with DFT calculations. (a-e) ARPES spectrum along the M-Γ-M high symmetry direction for (a) 3ML, (b) 4ML, (c) 5ML, (d) 7ML, and (e) 10ML PdTe$_2$ films. (f-j) DFT band structures for (f) 3ML, (g) 4ML, (h) 5ML, (i) 7ML, and (j) 10ML PdTe$_2$ films. The red boxes correspond to the range of the ARPES images above.

To understand the thickness dependence of the band structure, we took ARPES spectrum along $\bar{M}$-$\bar{\Gamma}$-$\bar{M}$ for 3ML, 4ML, 5ML, 7ML, and 10ML films (Figs. 3a-e respectively). The corresponding DFT calculations along the same direction are shown in (Figs. 3f-j). Apparently, the hybridization gap in the topological surface bands shrinks quickly as the film thickness increases, and vanishes when the thickness is larger than 5ML. This is because the coupling between surface states on opposite faces decay exponentially with the increase of film thickness. Through parabolic fitting the experimental bands (Figure 4a, b), we found that the gap in the topological surface bands is 0.24 eV and 0.13 eV for 3ML and 4ML, respectively. The experimental value of the hybridization gap is larger than the theoretical gap size (0.14 eV and 0.05 eV for 3ML and 4ML, respectively) as shown in the figure insets, which is because the DFT calculations typically underestimate the band gaps. To resolve these energy gaps more quantitatively, we plotted the energy distribution curve (EDC) at $\bar{\Gamma}$ and fit it with a Lorentzian multi-peak curve. We extracted the gap size from the Lorentzian fitting, see Figs. 4c,d. The peaks resolved in this analysis revealed the real experimental band gap sizes for the Dirac cone at -1.6 eV, however the resonance gap at -1.02eV is unable to be resolved due to the limited energy resolution of our ARPES experiments ~40 meV. Through the Lorentzian fitting we also found the gap in the topological surface bands is 0.06 eV for 5ML and 0 for 6ML and beyond. The film thicker than 6ML can be considered in the bulk limit under which there is no coupling between the opposite surfaces. The gap size at different thicknesses is summarized in Fig. 4e. We found a monotonic decrease in the gap size as the film thickness increases. The result indicates that the topological surface states of $PdTe_2$ can be engineered to have a variable gap size at the nodal point by controlling the film thickness.

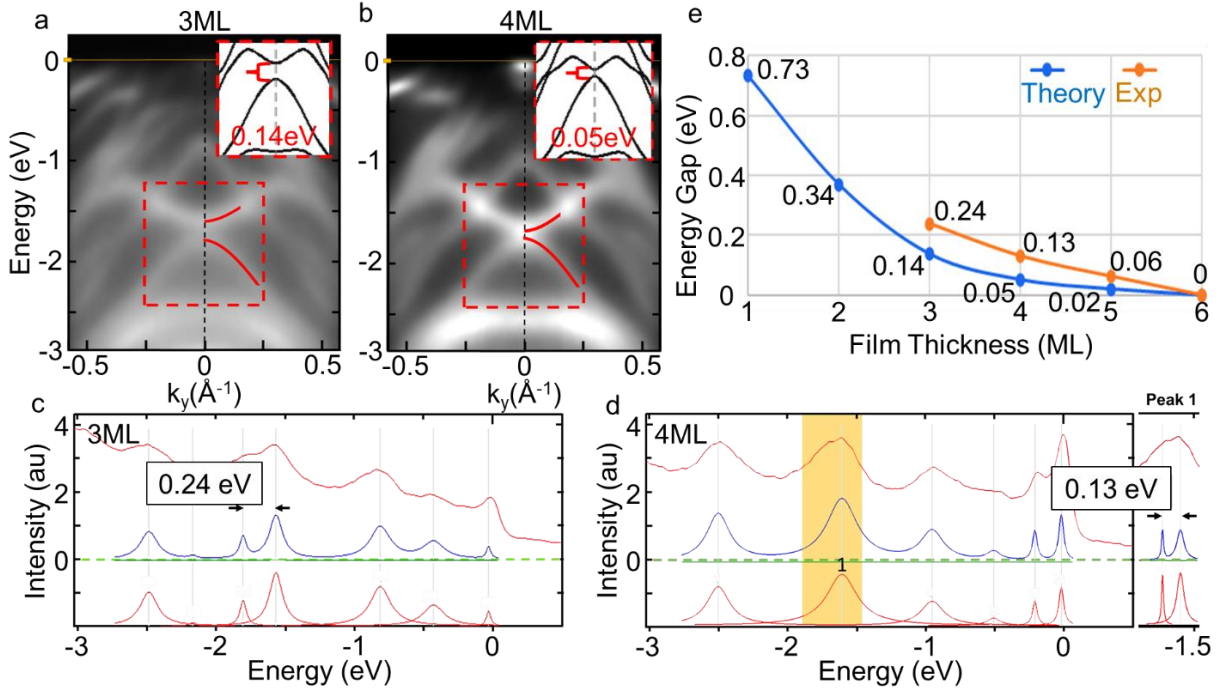

**Figure 4.** Band-gap Analysis of PdTe$_2$ films. (a, b) ARPES data along M-Γ-M for 3ML and 4ML PdTe$_2$ films, respectively. A parabolic fitting is shown for the Dirac cone at -1.67 eV. The inset shows the DFT bands. (c-d) ARPES intensity at Γ with Lorentzian peak fitting for (c) 3ML and (d) 4ML PdTe$_2$ films, respectively. The highlighted regions correspond to the zoom-in fittings to the right. (e) Thickness dependence of experimental energy gap of the topological Dirac cone compared to the DFT results.

The behavior of $PdTe_2$ growths on BLG/SiC was first shown in the work by E. Li *et al.*[29], where full films was reported to form for the thickness larger than 8ML. In our work, by optimizing growth methods (growth temperature, flux ratio, *etc.*), we demonstrated that high quality, ultrathin $PdTe_2$ films can grow and cover a majority of the surface (70%) which provide excellent STM and ARPES results that we can use to study the 2D limit properties of $PdTe_2$. We also detected the band dispersion for $PdTe_2$ quantum films below 6 ML by ARPES for the first time. For bulk $iPdTe_2$, the surface resonance band is buried in the bulk valence band. Approaching the 2D limit removes the majority of the bulk states and isolates this surface resonance band. This Rashba resonance persists in films down to 3ML thickness. Our ARPES spectra taken from $PdTe_2$ quantum thin films (in which the film thickness is comparable to the de Broglie wavelength of electrons) unambiguously demonstrates the isolation of the spin-split surface resonance bands from the bulk quantum well sub-bands. With the enhanced surface/bulk ratio, the spin-split surface resonance bands in ultrathin films are promising for spintronic applications.

More interestingly, we observed the hybridization band gaps in the topological surface bands in $PdTe_2$ thin films. The hybridization gap gradually diminishes as the film thickness increases, and vanishes for films thicker than 6ML. Therefore, 6ML can be considered as the critical thickness to distinguish the 2D and bulk behaviors of $PdTe_2$ films. It is known that $PdTe_2$ is a superconductor even with thickness down to 4ML[32] with theoretically predicted superconductivity even down to 1ML thickness[28]. If the Fermi level can be shifted to the energy region of the TSS, for example, by electric gating, then the system will behave like a topological superconductor thanks to the pair of spin-polarized electron in the TSS[33]. The hybridization band gaps observed in $PdTe_2$ quantum films provide an extra knob for tuning the topological properties of the band structure. When the Fermi level is moved from the energy of TSS states into the hybridization gap, the system will transform from a topological superconductor to a trivial superconductor. Through fine tuning on the local gating voltage, it is possible to construct a junction of topological and trivial superconductors and establish localized Majorana bound states at the domain walls. In addition, the ultrathin film geometry of PdTe$_2$ explored in this work facilitate the assembly of the superconducting device with a tunable gating voltage. Therefore, the spectroscopic characterization on the spin polarized surface resonance states and TSSs presented in this work establish PdTe$_2$ thin films an ideal platform for novel spintronic and topological quantum computation applications.